# NEURAL NETWORK APPROACH TO RAILWAY STAND LATERAL SKEW CONTROL

Peter Mark Beneš[1], Matouš Cejnek[1], Jan Kalivoda[2] and Ivo Bukovsky[1]

[1]Department of Instrumentation and Control Engineering, Czech Technical University, Prague, Czech Republic
`{PeterMark.Benes;Matous.Cejnek;Ivo.Bukovsky}@fs.cvut.cz`

[2]Department of Automobiles, Internal Combustion Engines and Railway Vehicles, Czech Technical University, Prague, Czech Republic
`Jan.Kalivoda@fs.cvut.cz`

*ABSTRACT*

*The paper presents a study of an adaptive approach to lateral skew control for an experimental railway stand. The preliminary experiments with the real experimental railway stand and simulations with its 3-D mechanical model, indicates difficulties of model-based control of the device. Thus, use of neural networks for identification and control of lateral skew shall be investigated. This paper focuses on real-data based modelling of the railway stand by various neural network models, i.e; linear neural unit and quadratic neural unit architectures. Furthermore, training methods of these neural architectures as such, real-time-recurrent-learning and a variation of back-propagation-through-time are examined, accompanied by a discussion of the produced experimental results.*

## 1. INTRODUCTION

An ongoing problem, which is currently under research in the railway industry, is that of lateral skew control of railway carriage wheel sets with Independently Rotating Wheels (IRW). In particular, the control of position of the leading wheel set of a railway carriage bogie, such to achieve central positioning of the wheel set, with respect to the span of the rail track. Or furthermore, to control the lateral deviation of the wheel set to follow a given desired set point. The primary necessity for such control arises from the need to improve behaviour of IRW wheel sets, as currently under study at CTU, on an experimental railway stand (roller rig). Particularly, for minimisation of wheel flange and rail head wearing as well as lateral forces and, furthermore, optimal stability at higher speeds of the wheel set. To date, various methods for control of the wheel set lateral position of IRW wheel sets have been under investigation, with both mechanical and electrical means of control [2], each featuring their own, individual drawbacks.

At CTU, the latest study features active control of the roller rig through manipulation of the yaw torque of the rig wheel set [1], via a state feedback and cascade PID control for a linearization of the model of the CTU roller rig. However, the results from this initial experimentation shows, that such method is not suitable for real time control. Thus, this paper aims to investigate the possible use of a neural network approach for lateral control of such railway wheel sets. The suitability of application to this problem is motivated by promising theoretical studies of higher-order neural units (HONUs), especially the quadratic neural unit for engineering problems for [3]-[7]. These studies are focused on the use of supervised learning based approaches for polynomial structured neural units, also known as a class of HONUs, for adaptive identification and control of real engineering systems. Further motivation arises from the successful implementation of a quadratic neural unit controller (Neuro-controller), for control of a bathyscaphe system located in the automatic control laboratories of CTU [12]. Where, here, such controller adhered more closely to the desired behaviour of the system than the conventionally used PID controller. An extension on this result can be recalled in the work [11]. Where, further study was made via introduction of a new software for adaptive identification and control, along with further testing on both a theoretical system and the previously mentioned bathyscaphe system. Given this, in this paper we aim to investigate use of neural network (NN) approaches in the following manner. We will begin by explaining in more depth the problem behind the recently employed state feedback and cascade PID control of the linearization of the CTU roller rig. The proceeding section will then describe the principles and control schemes behind the various experimented methods, for adaptive identification and control. Following this, an experimental analysis of the various approaches of NN, focusing firstly on adaptive identification of the CTU roller rig system and then to test the various NN methods for control. The final component of this paper will then be, to analyse and compare the produced experimental results, with a conclusion to be drawn at the end.

## 2. PROBLEM DESCRIPTION

This section aims to describe in more depth, the functionality behind the previously introduced experimental railway stand (CTU roller rig). More importantly, for the scope of this paper, we will elaborate on the necessity for control of such roller rig and the issues of conventional linear PID control or state feedback control, for controlling the lateral skew.

### 2.1 Experimental Setup of the Roller Rig at CTU

The below figure (Figure 1) depicts the experimental setup of the CTU roller rig, along with the 3-D model for design and simulation purposes. This rig features 5 motor drives, two pairs of 0.5m diameter rollers are independently driven, via the largest of the set of drives (FM1 & 2). To simulate both straight and curved tracks, similar to those present on real railway networks. A central servo motor (FM4) was introduced, to yaw the lower roller pair for replicating curved track motion. This setup however, assumes simulation of a rail pair without effects of rail buckling. For manipulation of the wheel set yaw, a separate servo motor, central to the wheel set (FM3) is installed. For the scope of this paper, this is the most crucial component, as it is the action of this servo motor that is used for control actuation for the lateral skew of the wheel sets. Where, the discussed control setups in this paper will analyse control via manipulation of this servo motor torque. Finally, a fifth drive FM5 is located between the front roller set, as a control differential.

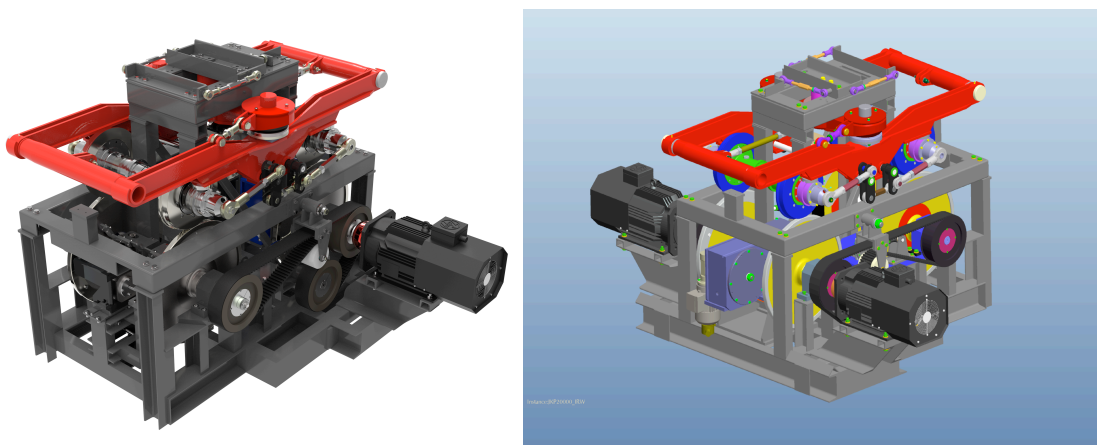

Figure 1: CTU Roller Rig (real left, PTC Pro/Engineer 3-D model on the right)

*2.2   Problem of the State Feedback and Cascade PID Control*

In the work [1], the method of lateral skew control via state feedback with cascade PID setup is introduced. In this presented setup, a linearization of the roller rig is used, via translation to a state space representation, achieved via a 12x12 matrix of dynamics. However, according to theory, this linearization of the roller rig system appeared to be uncontrollable and unobservable via analysis by common linear algebra and control approaches.

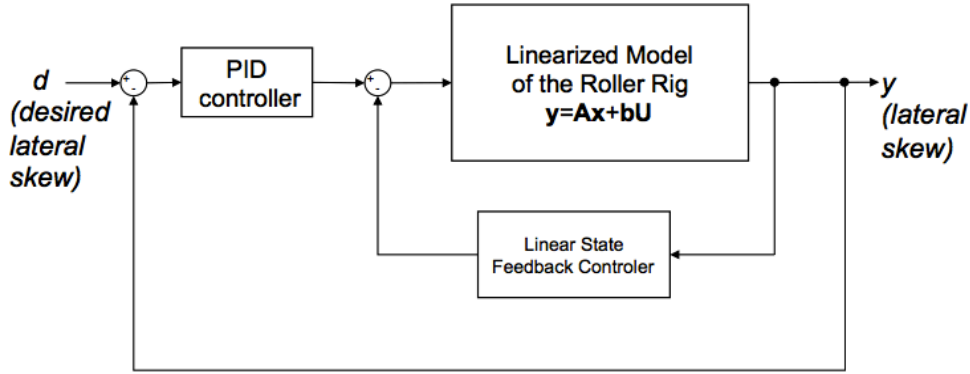

Figure 2: Control loop on linearization of the roller rig with State feedback + Cascade PID control

The above figure (Figure 2), depicts the control scheme for this setup. After numerous testing, it was found that a simulation step of approximately < 1E-6 was necessary, in order to achieve stability of the continuous control loop, which cannot be used for practical application. Hence, it is for this reason that an alternative approach is necessary for the lateral skew control of such railway wheel set. Thus, this paper aims on investigating the possibility of adaptive identification and control of the stand, based on measured data. For the purposes of this paper the data will be generated via a 3-D simulation model via software SIMPACK, linked via SIMAT toolbox with MATLAB Simulink, to provide real time simulation of the roller rig as a plant system for the investigated control approaches. Then, this data will be used as the training data for neural network models, to simplify otherwise complex simulation models and to allow a constant sampling for real time usability, furthermore, investigation into the potentials for real-time neural-network based control.

## 3.   APPLIED METHODS

In this section, the various NN approaches, aimed for adaptive identification and control of the previously introduced lateral skew problem, will be described. These approaches are based on the well know Gradient Descent (GD) method used as a tool for defining the learning rule of the applied neural units. These applied neural units (adaptive models) are trained via two methods, which will be focused on in this paper. These methods are namely, the method of incremental training or Real Time Recurrent Learning (RTRL) [8] applicable for dynamic adaptive models or a batch form of training, which is a variation of the back propagation through time or (BPTT) training [10], as an extension of RTRL training in combination with the famous Levenberg-Marquardt equation.

*3.1   Preliminaries*

This subsection aims to review the important theories from works [11]-[13], behind the GD method and structures of the neural units used within this paper. Firstly, we will begin by introducing the famous GD algorithm for the linear and quadratic neural units. To understand this, we must begin with the polynomial models of the linear and quadratic (LNU and QNU) neural units respectively, as follows

$$y = \sum_{i=0}^{n} x_i w_i = w_0.x_0 + w_1.x_1 + ... + w_n.x_n = \mathbf{w.x} \tag{1}$$

$$y = \sum_{i=0}^{n}\sum_{j=0}^{n} x_i x_j w_{i,j} = w_{0,0}x_0 x_0 + w_{0,1}x_0 x_1 + ... + w_{n,n}x_n^2 = \mathbf{rowx.colW} \tag{2}$$

Where, $y$ represents the output of the LNU (1) and QNU (2) respectively. In regards to the LNU, $\mathbf{w}$ stands for an updatable vector of neural weights and $\mathbf{x}$, represents the input values of the engineering system in the case of a purely static model and in the sense of a dynamic model, a combination of inputs of the real system and neural model outputs. Looking at equation (2), **rowx** is a long-vector representation of the utilised input vector. Where, **colW** represents a weight matrix of the quadratic neural unit in general. From this, we may then understand the GD algorithm as applied to such neural units.

$$w_{i+1} = w_i + \mu.e(k).\frac{\partial y(k)}{\partial w_i} \tag{3}$$

$$\mathbf{colW}(k+1) = \mathbf{colW}(k) + \mu.e(k).\frac{\partial y(k+1)}{\partial \mathbf{colW}} \tag{4}$$

Here, equation (3) & (4) depicts the GD algorithm for both the LNU and QNU respectively. Here the output of the GD algorithm is to incrementally over sample-by-sample, update the neural weights such to adaptively teach the LNU or QNU model, to learn the engineering system. Here, μ represents the learning rate of the GD algorithm. This is analogical to humans where, setting μ relatively high, corresponds to faster learning of the human and consequently, means the less detail the human can remember and digest from their learning. Furthermore, setting this parameter to a smaller value, corresponds to a slowing rate of learning i.e. the human may remember the information learned, quite well, but less information overall about the subject. The next parameter is $e(k)$ ($k$ representing the number of the sample), this represents the current error between the real and calculated output of the model. The final term $\frac{\partial y(k)}{\partial w_i}$, corresponds to the partial derivatives of the neural unit output, with respect to the individual neural weights.

Regarding the QNU in equation (4), we see that the GD algorithm is analogical, with exception of updating the matrix of neural weights as opposed to a vector in the sense of the LNU.

Till now, the structures of GD for LNU and QNU were reviewed in the format of RTRL or sample-by-sample method of learning. Where, the neural weights are updated over each new sample of the engineering system data. However for certain engineering processes, it is more advantageous to use the BPTT batch method of training these neural weights, over runs of the algorithm rather than over each sample. This is because the RTRL method focuses on the contemporary governing law of the system as opposed the BPTT which focuses more so on the main governing law of the input and outputs of the engineering system. The BPTT method is achieved via an extension of the RTRL method with combination of the famous Levenberg-Marquardt equation. It is also important to note that this method is more preferable in cases where the data may be affected by noise. The following equation denotes the weight update rule for the BPTT method;

$$\Delta \mathbf{w} = (\mathbf{J}^T.\mathbf{J} + \frac{1}{\mu}.\mathbf{I})^{-1}.\mathbf{J}^T.\mathbf{e} \qquad (5)$$

Here, the neural weights are updated over each run of the algorithm (or batch trained) in the following way w=w+Δw. Here equation (5) describes the change, necessary for the update of the batch trained weights. Where, **J** represents the Jacobian matrix of derivatives for the neural unit. This may be the complete partial derivatives of the neural model with respect to each neural weights, or in practical applications it seems useful to simply introduce this Jacobian matrix as the input vector or matrix itself, being **x** and **colx** for LNU and QNU respectively. Furthermore, it is important to note that $\mathbf{colx} = \mathbf{rowx}^T$.

Often in such adaptive neural units, it is apparent that a modification of the normalised learning rate may be used to solve issues associated with instability of learning. In practise, it is possible to employ the simplified normalised learning rate as presented in the work [11], as follows;

$$\eta = \frac{\mu}{\mathbf{x}(k)\mathbf{x}(k)^T + 1} \qquad (6)$$

Where, equation (6) represents the normalised learning rate in the sense of LNU. This is analogically represented in the QNU, by replacement of **x**(k) (i.e. each k th sample of the input vector x) with **colx**(k). It should be noted that the above representation holds for RTRL training for dynamic adaptive models, where the algorithms used in this paper for BPTT method of training, take the learning rate μ itself.

### 3.2 Adaptive Identification and control

The previous section focused on usage of neural units in the sense of adaptive identification of a real engineering system. In this subsection, we extend on these neural units as a method of control, with brief review from works [11]-[13], and extension as applied to the lateral skew problem, which is indeed our focus within this paper.

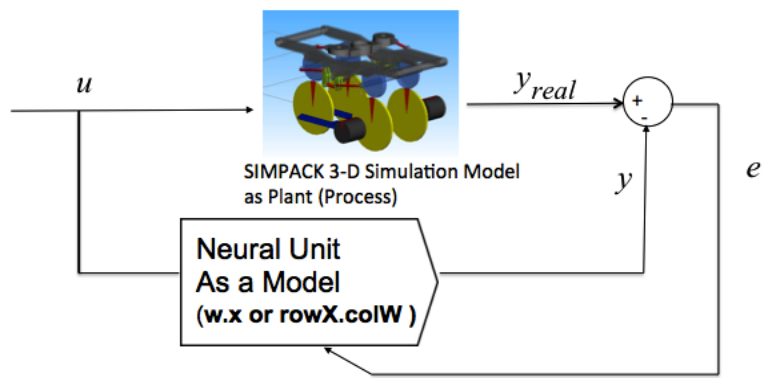

Figure 3: Adaptive Identification with supervised learning of neural networks (where, **w.x**=LNU, **rowX.colW**=QNU)

The above figure (Figure 3), depicts the identification scheme for the previously reviewed neural network architectures. Where, for the scope of this paper a simulation of the real roller rig will be used for data generation of the investigated neural network approaches for control. *u,* represents the input data of the roller rig, in our case this is the yaw torque of the servo motor system for manipulating the lateral skew of the wheel set. The output $y_{real}$, is the simulated output from the real time 3-D SIMPACK model and $y$, being the output of the neural unit, and the difference being the error, $e$.

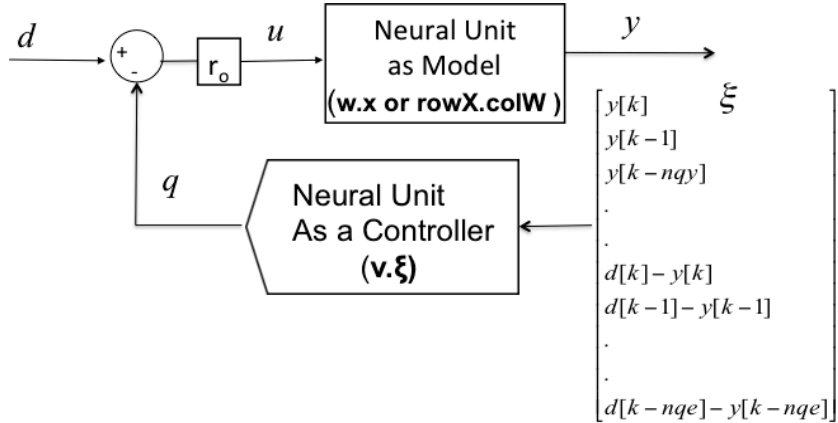

Figure 4: Adaptive control loop for experimental study of a neural network controller (where, **w.x**=LNU, **rowX.colW**=QNU)

The above (Figure 4), shows an extension of the discussed neural architectures for application to lateral skew control of the roller rig system. Here, once the neural unit as a model is identified, it may be utilised as the foundation for a neural network based control setup. For the scope of this paper, we will propose an offline tuned control scheme, as the goal here, is to investigate the potentials of applying a neural network based control, for this application. However, the extension to online control is indeed the ultimate aim of our research to this problem, beyond this paper. Figure 4, depicts use of a second neural unit as a controller. Similarly to the previously mentioned architectures, as a controller here too the neural unit may take shape to that of the LNU or QNU adaptive models. However, in this case the adaptive neural weights are tuned differently to that of those used for the neural unit as a model and hence, depicted as $v$. Analogically, should a QNU structure be applied, these neural weights would be represented in a matrix form. Further to this, the above figure (Figure 4) introduced a new vector $\xi$. This vector comprises of a combination of outputs from the neural unit as a model and the difference between the desired behaviour (in our case desired yaw torque or lateral skew of the roller rig) and the output of the neural model. **v. $\xi$** or collectively, the variable $q$, thus serves as a manipulator for the newly feed samples of the neural unit as an identified model. Here the GD algorithms are employed in the following manner to achieve sample-by-sample adaptation of the neural weights for the controller, as follows;

$$v_{i+1} = v_i + \mu.ereg(k).\frac{\partial y(k)}{\partial v_i} \quad (7)$$

Where $v_i$, are adaptable neural weights of the neural unit as a controller and $e_{reg}(k)$ is the error between the desired value of the real system (in our case the roller rig, where the desired value will be denoted as $d$) and the real system output value at sample $k$. $\frac{\partial y(k)}{\partial v_i}$, is the partial derivative of the output of the neural unit as a model, with respect to the individual adaptive neural weights of the neural unit as a controller. An extension of this weight update scheme for BPTT training would result in the following form $v=v+\Delta v$, where the change of neural weights for each batch would be analogical to equation (5).

# 4. EXPERIMENTAL ANALYSIS

This section aims to analyse the previously reviewed methods of adaptive identification and control, via the introduced neural network architectures. Particularly, we aim to investigate the applicability of the discussed neural architectures for this problem of the lateral skew control of railway wheel sets. In this section, we will simulate our results using simulation data of the previously described 3-D SIMPACK model (at mid-range speed) linked with MATLAB Simulink in real time simulation. The data was produced with 0.001 sampling over a 10 second interval. First, we will analyse the results for identification of the roller rig as a system, via the different methods of the neural network models, followed by an extension of control, where the various combinations of neural network architectures will be tested.

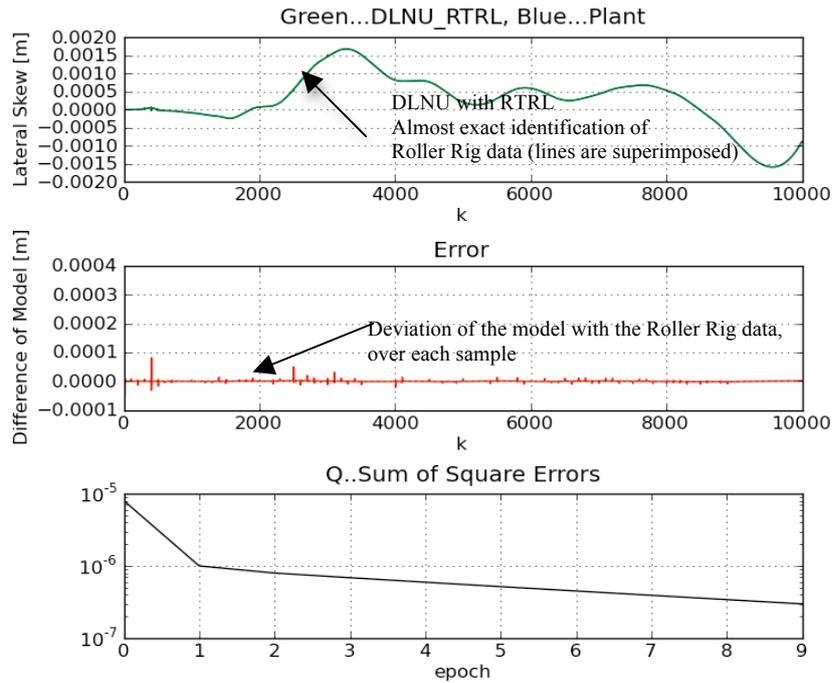

Figure 5: Testing of the adaptive identification, where the plant (roller rig) is represented by blue colour and green being the neural model. Trained by DLNU with RTRL Training; μ=1, epochs =10, For *ny*=3 (previous samples of model output) and *nu*=5 (previous samples of real system input)

The above figure (Figure 5) and following figure (Figure 6), illustrates the adaptive identification process of the roller rig system data. We can note, that both methods of the dynamic LNU or (DLNU) and dynamic QNU or (DQNU), for RTRL learning methods, achieved almost exact identification. Where, the DQNU performed slightly faster in terms of convergence to minima of sum of square errors, as opposed to the DLNU.

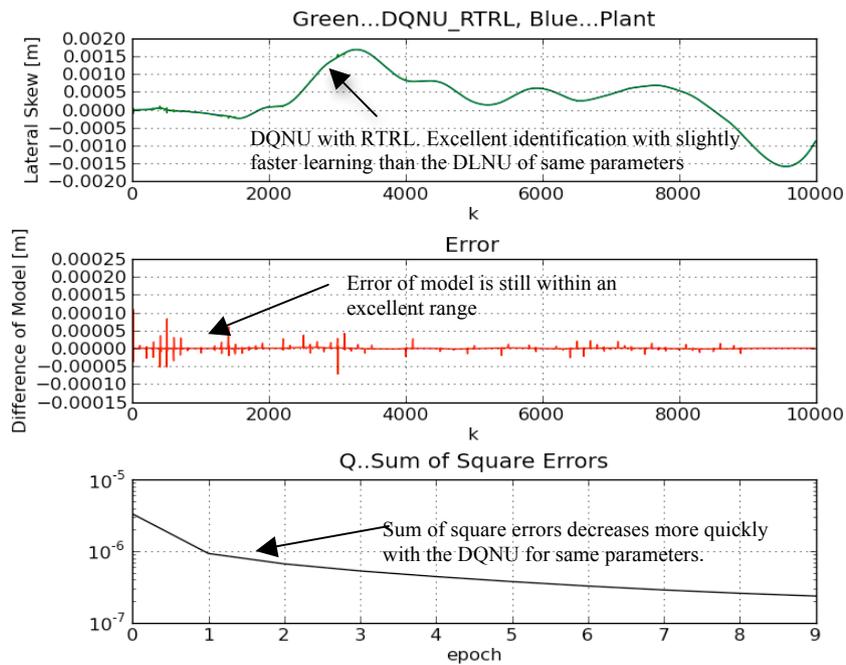

Figure 6 : Testing of the adaptive identification, where the plant (roller rig) is represented by blue colour and green being the neural model. Trained by DQNU with RTRL Training; μ=1, epochs =10, For *ny*=3 (previous samples of model output) and *nu*=5 (previous samples of real system input)

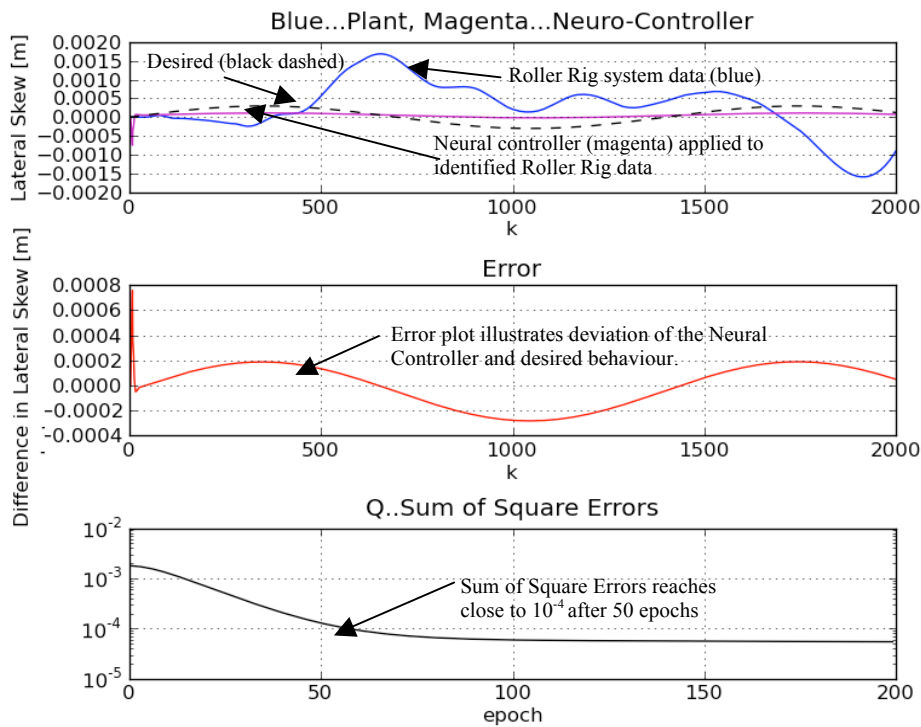

Figure 7: Testing of the adaptively tuned control loop, where the roller rig is represented by trained DLNU with constant and previously trained parameters (Figure 5) and the adaptive feedback controller is trained via LNU, after 10 epochs Adaptive Identification (of DLNU with RTRL) + Neural Unit Controller for lateral skew control of roller rig system - Using LNU with Incremental training μ=0.1, Data re-sampling = each $5^{th}$ sample, epochs =200, for *nqy*=0 (previous samples of neural model output) and *nqe*=2 (previous samples of difference of the neural model output and desired behaviour of the system.

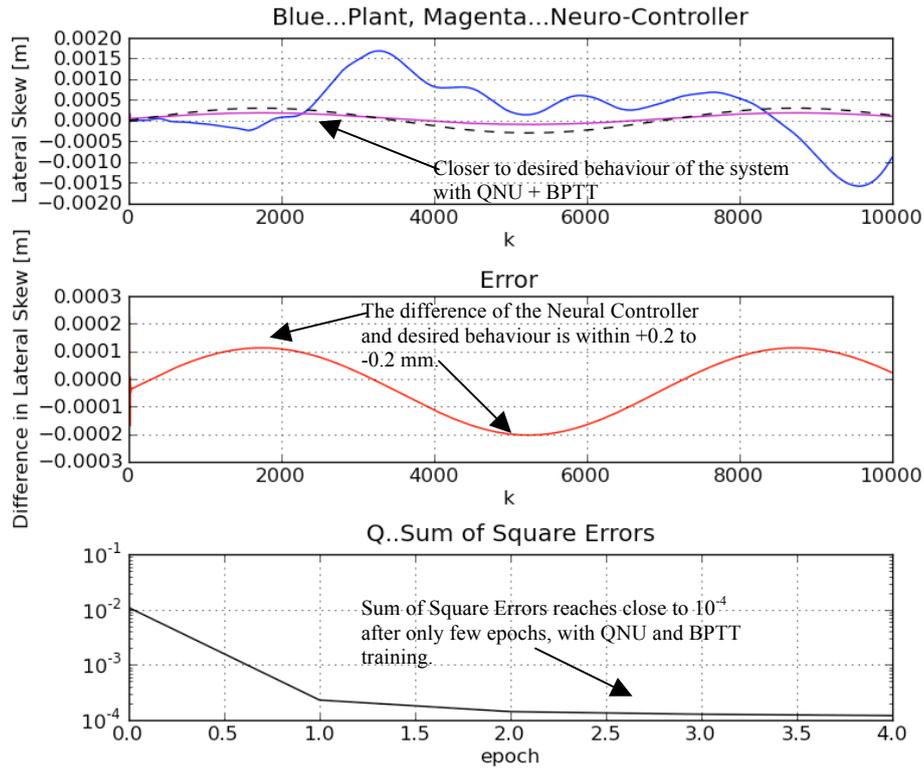

Figure 8: Testing of the adaptively tuned control loop, where the roller rig is represented by trained DQNU with constant and previously trained parameters (Figure 6) and the adaptive feedback controller is trained via QNU,
after 10 epochs Adaptive Identification (of DQNU with RTRL) + Neural Unit Controller for lateral skew control of roller rig system - Using QNU with BPTT Training μ=0.00818, epochs=5, for *nqy*=0 and *nqe*=3

Here, Figure 7 & Figure 8 show the application of the various neural units for control of the lateral skew. The black dotted line, illustrates the chosen desired behaviour of the roller rig lateral skew. It was found that the BPTT training method used in combination with the QNU had the best performance. Providing closer control to the desired behaviour, as compared with the incremental training methods of both LNU and QNU architectures.

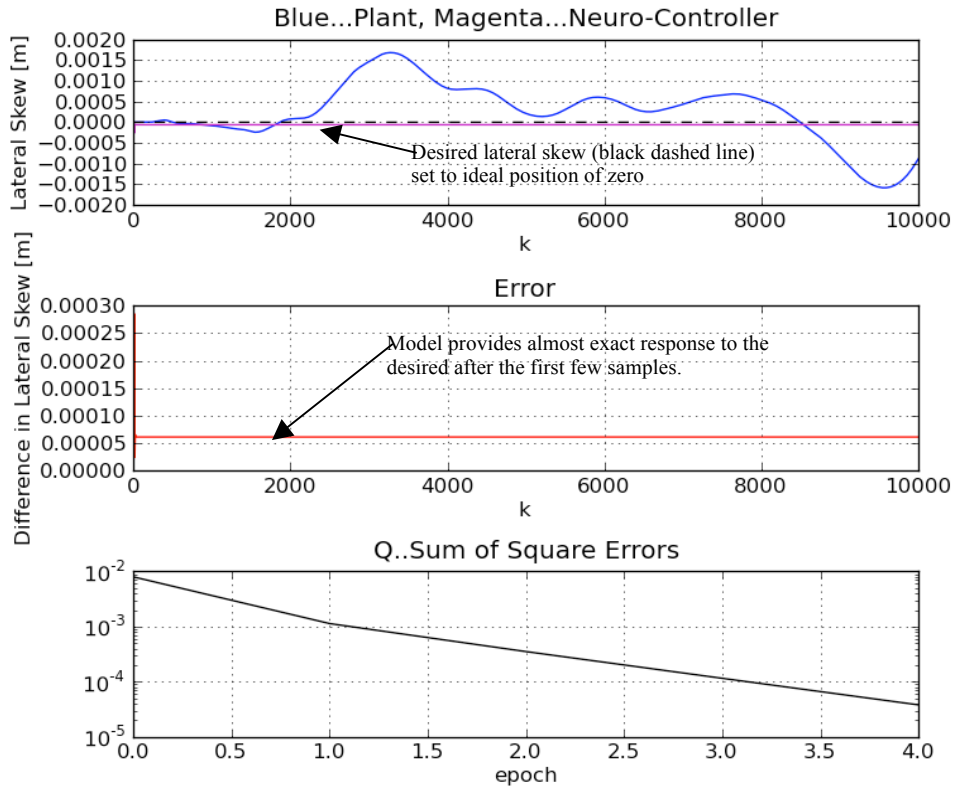

Figure 9: Testing of the adaptively tuned control loop, where the roller rig is represented by trained DQNU with constant and previously trained parameters (Figure 6) and the adaptive feedback controller is trained via QNU,
after 10 epochs Adaptive Identification (of DQNU with RTRL) + Neural Unit Controller for lateral skew control of roller rig system - Using QNU with BPTT Training μ=0.001, epochs=5, for *nqy*=3, *nqe*=3

Figure 9 depicts a theoretical test of the roller rig system under an idealised case of zero lateral skew. This is included in our results to investigate the capability of the used neural controller. From this figure, we see that after several small samples at the beginning of the controller's application the controller provides almost exactly zero lateral skew. This result however, must be reasoned with what is really possible in terms of actuation of the real system and real conditions of the rig, such to achieve this ideal of a result.

## 5. DISCUSSION OF EXPERIMENTAL RESULTS

In the previous section, the experimental results of the various neural network architectures for adaptive identification and control were presented. From the first set of figures, regarding the adaptive identification for the roller rig system, it was found that the most suitable learning method for both the DLNU and DQNU was the RTRL training method. Here, relatively few epochs were adequate for both architectures to identify the system data, with almost exact identification. The DQNU model was slightly faster in learning the behaviour of the system, thus, was used as the identified neural unit as a model, in the control section. The extension of these various architectures for control showed more distinguishable results. The application of both LNU and QNU with incremental learning for the neural controller was only possible when a higher number of epochs were run, or higher learning of the model with a small learning rate.

Few runs of the algorithm were unable to match the desired behaviour even moderately. However, after the first few samples of application as in Figure 7, reasonable behaviour of the controller was achieved. The combination of identification via the DQNU followed by extension of a QNU trained by BPTT on the neural controller, showed to follow more closely to the desired behaviour of the roller rig system than the other tested combinations. Where, the setup of the QNU with BPTT neural controller featured input vector $\xi$, being comprised only of several errors or differences of the desired behaviour, with the output of the neural unit as a model. Further to this, the training was substantially quicker than the incremental training method, showing that only several epochs were sufficient for the model to achieve its optimal behaviour with respect to the desired of the system.

In the above results, two forms of the desired behaviour were used. The first was under proposal that the lateral derivation of the wheel set would be within a small tolerance range, which is most realistic in real application of such control. The presented desired data is relative to the small range of lateral skew in the simulation data. Where, within this exampled range, we could be sure the wheel flanges would not contact the rails of the real roller rig. The desired behaviour is however relative to the setup of the wheel sets and rails, thus the desired deviation and lateral skew outputs may indeed vary for real railway applications. The second was an idealised situation where no lateral skew at all would occur. This is principally unrealistic for the neural controller to achieve zero lateral skew, due to a combination of natural factors on real railway wheel sets, however for demonstrational purposes, we may note the capability of the neural controller, providing almost exact behaviour of the idealised lateral skew, as applied on the identified roller rig system.

## 6. CONCLUSION

Referring back to the original control loop depicted in Figure 1 (as presented in work [1]), of the Simulink model of the linearization of the roller rig system and state feedback with cascade PID controller. We can recall that a minimum sampling of <1E-6 seconds for numerical stability was necessary, which is not possible for practical application. After investigation of the presented neural network approaches, we may conclude that it has promising potential for real application on controlling the lateral skew of IRW railway wheel sets. Now, we note that the achieved sampling can be within order of 0.005 (5E-3) seconds, for adequate functionality of the neural network based, adaptive identification and control system as applied to the roller rig. The tested models in this paper were DLNU and DQNU for adaptive identification models and LNU or QNU architectures as a controller. The experimental results of these architectures showed that both DLNU and DQNU, precisely approximated the complex SIMPACK model of the roller rig system. This result however could vary when using more or, real training data measured from the rig itself. In terms of control of the lateral skew, the QNU architecture showed better results to keep the lateral skew within close limits to the desired behaviour of the system, which is also in general consensus with our previous findings on using QNU for control. We also can note the significantly faster tuning of the control loop via QNU as compared to the LNU, more particularly, the BPTT training method as applied to the controller compared to the incremental method. Where, more desirable control was shown, in much fewer epochs or runs of the controller algorithm. Thus, this case study shows that with proper tuning of the investigated neural units, application to the real system is theoretically possible, to achieve adequate control for this investigated problem. Thus, real application of such adaptive control loop of neural network based architecture, is indeed the very next step of our research.

## ACKNOWLEDGEMENTS

The authors would like to acknowledge the following grant for support during this work: SGS12/177/OHK2/3T/12SGS12/177/OHK2/3T/12, Non-conventional and Cognitive Signal Processing Methods of Dynamic Systems.
And also the Technology Agency of Czech Republic. Project No: TE01020038 "Competence Center of Railway Vehicles".

## Short Biography of the Authors

**Peter Mark Beneš** received his Bachelor's degree with honours at Czech Technical University (CTU) in Prague in 2012. Currently he is a Master's student, with expected PhD studies to follow in 2014. His research focuses on non-conventional neural networks for adaptive identification and control of industrial systems including hoist mechanisms and skew control of rail-based mechanisms as such cranes and railway vehicles. Peter's work has been awarded in local and international student competitions also with an industrial BOSCH award in 2013.

**Matouš Cejnek** received his Bachelor's degree at Czech Technical University (CTU) in Prague in 2012. Currently he is a Master's student at Czech Technical University in Prague. His research focuses on non-conventional neural networks for adaptive systems and novelty detection in time series and biomedical applications. Matous's work has been awarded in local and international student competitions 2013.

**Jan Kalivoda** graduated from Czech Technical University (CTU) in Prague where he received his Master's degree with honours in 1996 and Ph.D. in the field of Machines and Equipment for Transportation in 2006. Currently he is a teacher and active researcher in the Department of Automotive, Combustion Engine and Railway Engineering at CTU. His research interests include MBS models of railway vehicles, Mechatronics for railway vehicles, Active control of railway vehicle suspensions and wheel sets.

**Ivo Bukovsky** graduated from Czech Technical University in Prague (CTU) where he received his Ph.D. in the field of Control and System Engineering in 2007 and became associate professor since 2013. His research interests include higher-order neural networks, adaptive evaluation of time series and systems, multi-scale-analysis approaches, control and biomedical applications. He was a visiting researcher at the University of Saskatchewan (2003), at the University of Manitoba in Canada (2010), and he was a visiting professor at Tohoku University (2011)